\documentclass[11pt,tightenlines,onecolumn, aps, prd, nofootinbib, superscriptaddress, showkeys, showpacs]{revtex4-1}

\usepackage{latexsym}
\usepackage{amssymb}
\usepackage[dvips]{graphicx}
\usepackage{color}
\usepackage{graphicx}

\usepackage{amsmath, amsfonts, amssymb, mathrsfs}
\usepackage{amsthm}
\usepackage{tensor}
\usepackage{multirow}
\usepackage{bbm}
\usepackage[geometry]{ifsym}

\newcommand{\Dslash}{{D\kern-0.63em{/}}}

\usepackage{graphicx}
\usepackage{float}
\usepackage[font=small]{caption}

\begin{document}

\newcommand{\beq}{\begin{equation}}
\newcommand{\eeq}{\end{equation}}

\title{Beyond the Equivalence Principle: Gravitational Magnetic Monopoles}

\author{M. Novello}

\affiliation{ Centro de Estudos Avan\c{c}ados de Cosmologia (CEAC/CBPF) \\
 Rua Dr. Xavier Sigaud, 150, CEP 22290-180, Rio de Janeiro, Brazil. }

\author{A.~E.~S. Hartmann}
\affiliation{Dipartamento di Scienza e Alta Tecnologia, Universit\`{a} degli Studi dell'{}Insubria,  \\
 Via Valleggio 11, 22100 Como, and INFN Sez di Milano, Italy.}

\date{\today}

\begin{abstract}
We review the hypothesis of the existence of gravitational magnetic monopoles (H-pole for short) defined in analogy with the Dirac\rq s hypothesis of magnetic monopoles in electrodynamics. These hypothetical dual particles violate the equivalence principle and are accelerated by a gravitational field. We propose an expression for the gravitational force exerted upon an H-pole. According to GR ordinary matter (which we call E-poles) follows geodesics in a background metric $g_{\mu\nu}.$ The dual H-poles follows geodesics in an effective metric $\hat{g}_{\mu\nu}.$
\end{abstract}

\vskip2pc
 \maketitle

\section{Introduction}
Einstein\rq s General Relativity describes the effect of gravity on any particle as the modification of the background geometry. Particles, free of any other interaction, follow geodesics in this modified metric. In other words, gravity does not accelerate bodies. Neverthless we will analyze here a suggestion \cite{novelloetal}, \cite{salim} according to which there may exist particles (as yet not observed) such that gravity could be responsible for its acceleration. How is this possible? In order to examine such hypothesis we follow the original idea of existence of magnetic monopoles made by Dirac long time ago \cite{dirac}.

The gravitational correspondence come from the observation that the connecting vector of neighbouring geodesics (which we call E-pole) is controlled by the electric  part $ E_{\mu\nu}$ of Weyl tensor in a space free of matter. This led us by an analogy with Dirac\rq s procedure to suggest that there could be possible to consider the existence of paths followed by real particles (which we call H-pole) such that its connecting vector is controlled by the magnetic part of Weyl tensor ($ H_{\mu\nu}$) (see the appendix for a short compilation of Weyl tensor and notation).

%The standard paths, the geodesics, is called an electric gravitational monopole (E-pole). The curves accelerated by a gravitational field  will be called a magnetic gravitational monopole ( H-pole).

 In the next sections we describe shortly the original idea of electromagnetic magnetic monopole and the analogous idea for the gravitational magnetic monopole and suggest the formula for the force on an H-pole.

\section{Dirac magnetic monopole}

The equation of motion of a charged particle with charge $ e$ and mass $m$ in a given electromagnetic field is given by
\begin{equation}
\frac{d^{2}x^{\mu}}{ds^{2}} = \frac{e}{m} \, F^{\mu\nu} \, \frac{dx_{\nu}}{ds} = \frac{e}{m} \, \, E^{\mu},
\label{17nov1}
\end{equation}
that is, an electric monopole couples only with the electric part of the electromagnetic field.

Dirac \cite{dirac} made the hypothesis, based on dual properties, that it could exist a magnetic monopole that obeys the dual equation
\begin{equation}
\frac{d^{2}z_{\mu}}{ds^{2}} = \frac{g}{m} \, F^{*}_{\mu\nu} \, \frac{dz^{\nu}}{ds} = \frac{g}{m} \, \, H_{\mu}.
\label{17nov2}
\end{equation}

Let us consider now the motion of particles in a gravitational field. Before let us make a brief comment on Jacobi field.

\subsection{Jacobi field}

A vector $ Z^{\alpha}$ which connects points on two infinitesimally neighboring geodesic curves $ \Gamma$ of a congruence $ x^{\alpha}(s) $ with equal values of the parameter $ s $ is called a connecting vector. For a Ricci flat geometry ($ R_{\mu\nu} = 0$) we call a Jacobi field along $ \Gamma$ any connecting vector $ Z^{\alpha}$ that satisfies the equation
\begin{equation}
\frac{D^2}{Ds^2}\ Z^\alpha  + W^{\alpha}{}_{\mu\beta\nu} \, v^{\nu} \, v^{\mu} \, Z^{\beta} = 0,	
\label{14Fev21}
\end{equation}
which in terms of the electric part of the Weyl tensor yields (see the Appendix for the definitions)
\begin{equation}
\frac{D^2}{Ds^2}  Z^\alpha  = - E^{\alpha}{}_{\beta} \, Z^{\beta}.	
\label{15Fev21}
\end{equation}
Let us note that in the case of spin one the electric (magnetic) part of the field is obtained by a single projection of the field upon the velocity vector of an observer. For the case of spin two, the corresponding electric (magnetic) part needs a double projection, as we shall see.

For accelerated curves, the generalized Jacobi field is defined \cite{novellodamiao} in terms of a polynomial function $\tensor{N}{^\alpha_\beta}$ of the curvature tensor satisfying the equation 
\begin{equation}
\frac{D^2}{Ds^2}  Z^\alpha  = \tensor{N}{^\alpha_\beta} \, Z^{\beta}.	
\label{13mar21}
\end{equation}

\section{Gravitational Magnetic Monopole}

In the General Relativity theory a test body follows a geodesic in a given metric. The important point of contact with the dual framework pointed out by Dirac comes from the remark that the evolution of the connecting vector of two neighbouring particles, with the same value of the affine parameter, call it $ \eta^{\alpha}$ satisfies Jacobi equation (\ref{15Fev21}). This is the motivation to call these ordinary matter as E-poles.

Following the dual approach let us make the hypothesis that there exists particles the path of which is not a geodesic but that is accelerated by gravity. This will mimic the procedure that led to the suggestion of magnetic monopoles \cite{dirac}.

The dual operation led us to construct a congruence of accelerated curves in a gravitational field such that its corresponding connecting vector $ \Pi^{\alpha}$ satisfies the dual equation that generalizes Jacobi equation:
\begin{equation}
D^{2} \Pi^\alpha / D s^{2} = - \, {^{*}W}^{\alpha}{}_{\mu\beta\nu} \, v^{\nu} \, v^{\mu}  \pi^{\beta} =  - \, H^{\alpha}{}_{\beta} \, \Pi^{\beta}.
\label{15220}
\end{equation}

The gravitational force produces an accelerated path on an H-pole  (see \cite{novelloetal} for details)

\begin{equation}
\frac{d^{2}}{ds^{2}} z^{\alpha} (s) + \Gamma^{\alpha}_{\mu\nu} \, \frac{d z^{\mu}}{ds} \frac{d z^{\nu}}{ds} = F^{\alpha}
\label{1850}
\end{equation}

where $F^{\alpha}$  satisfies the condition
\begin{equation}
\left( F_{\alpha ; \mu} + R_{\alpha\beta\mu\nu} \, v^{\beta} \, v^{\nu} + H_{\alpha\mu}\right) \, \Pi^{\alpha} = 0.
\label{11setembro}
\end{equation}
In the particular case in which
\begin{equation}
 F_{\alpha ; \mu}=  - \, H_{\alpha\mu}- R_{\alpha\beta\mu\nu} \, v^{\beta} \, v^{\nu}
\label{11setembro2}
\end{equation}
it satisfies the two conditions:
\begin{itemize}
\item{ $ F_\alpha$ is a gradient $F_{\alpha} = d \Phi / dx^{\alpha}$}
\item{ The scalar field $\Phi$ obeys the wave equation $ \Box \Phi = - R_{\mu\nu}  \, v^{\mu} \, v^{\nu}.$}
\end{itemize}

In the case the Ricci tensor ($ R_{\mu\nu}$) does not vanishes the expression (\ref{15220}) must be modified once there are two possibilities of taking the dual of the Riemann curvature tensor:
$$*R^{\alpha\beta\mu\nu} := \frac{1}{2} \, \eta^{\alpha\beta\rho\sigma} \, R_{\rho\sigma}{}^{\mu\nu}, $$
and
$$ R* ^{\alpha\beta\mu\nu} := \frac{1}{2} \, \eta^{\mu\nu\rho\sigma} \, R_{\alpha\beta}{}_{\rho\sigma} .$$

Thus we set

$$
D^{2} \Pi^\alpha / D s^{2} + \frac{1}{2} \, \left( *R^{\alpha}{}_{\mu\beta\nu} + R* ^{\alpha}{}_{\mu\beta\nu} \right)  \, v^{\nu} \, v^{\mu}  \pi^{\beta} = 0.$$
This modification does not changes the properties of the force, that still satisfies the conditions that $ F_\alpha = d \Phi / dx^{\alpha}$ and the scalar field $\Phi$ obeys the wave equation $ \Box \Phi = - R_{\mu\nu}  \, v^{\mu} \, v^{\nu}.$

\subsection{The gravitational force}

The very important result that it is possible to annihilate the local gravitational force was the true basis of the program of geometrization done by Einstein in the theory of general relativity (GR). It implies that the net effect of the gravitational field on a body is to produce the geodesics that become the natural paths that all bodies free from any other force must follow. In Maxwell Electrodynamics the acceleration of a charged body of mass $ m,$ electric charge $ q$ and velocity $ v^{\mu} $ that an electromagnetic field induces is given by Lorentz formula

\begin{equation}
a_{\mu} = \frac{q}{m} \, F_{\mu\nu} \, v^{\nu}.
\label{16novembro1}
\end{equation}

The hypothesis of the existence of a gravitational magnetic pole (H-pole) imply that it does not follow geodesics. Thus the acceleration $ a^{\mu}$ of an H-pole in the geometric framework must depend only on the geometric properties.

Besides, once the natural motion of E-poles does not contain any specific constant related to the body (that is, the equivalence between inertial and gravitational masses) it is natural to make the hypothesis that the same must occurs for an H-pole. The question is to find an expression for $ a^{\mu} $ that has such property. There is no better way than to make an appeal to the potential of Weyl conformal tensor, that is the Lanczos tensor. In other words, we will make the hypothesis that in the case in which the Ricci tensor vanishes the acceleration of an H-pole is given in terms of the Lanczos tensor.

\subsection{Lanczos tensor}

Let us consider the third order tensor $ L_{\alpha\beta\mu}$ introduced by Lanczos \cite{lanczos}, that obeys the following relations
\begin{align}
 L_{\alpha\beta\mu} + L_{\beta\alpha\mu} &= 0,
 \label{17nov4} \\[1.5ex]
 L_{\alpha\beta\mu} + L_{\beta\mu\alpha} + L_{\mu\alpha\beta} &= 0.
 \label{17nov5}
\end{align}
The conditions (\ref{17nov4})-(\ref{17nov5}) implies that such tensor has only $20 $ independent components. Lanczos showed \cite{lanczos} (see  also \cite{bampicaviglia}) that Weyl conformal tensor can be written in terms of $L_{\alpha\beta\mu}$ as follows:

\begin{equation}
 W_{\alpha\beta\mu\nu} =  L_{\alpha\beta[\mu;\nu]} + L_{\mu\nu[\alpha;\beta]} + S_{{\alpha\beta\mu\nu}} + \frac{2}{3} \, L^{\sigma\lambda}{}_{\sigma ; \lambda} \, g_{\alpha\beta\mu\nu}
 \label{17nov6}
\end{equation}
where
$$ S_{\alpha\beta\mu\nu} := \frac{1}{2} \, \left(L_{(\alpha\nu)} \, g_{\beta\mu} +  L_{(\beta\mu)} \, g_{\alpha\nu} -  L_{(\alpha\mu)} \, g_{\beta\nu} - L_{(\beta\nu)} \, g_{\alpha\mu} \right),$$
$$ L_{\alpha\mu} := L_{\alpha}{}^{\sigma}{}_{\mu ; \sigma} - L_{\alpha}{}^{\sigma}{}_{\sigma ; \mu} $$
$$ g_{\alpha\beta\mu\nu} :=  g_{\alpha\mu} \,  g_{\beta\nu} - g_{\alpha\nu} \, g_{\beta\mu}.$$
We are using $ (a b) := ab + ba$ and $ [a b] := ab - ba.$

A very important result \cite{novellovelloso}  comes from the remark that it is possible to use the irreducible components associated to a congruence of curves to write the Lanczos tensor \cite{bonilla}. We then set

\begin{equation}
L_{\alpha\beta\mu} = \sigma_{\mu [\alpha} \,  v_{\beta]} + \omega_{\alpha\beta}\, v_{\mu} - \frac{1}{2} w_{\mu [\alpha} \, v_{\beta]} +  \frac{3}{2} \, a_{[\alpha} \, v_{\beta]} \, v_{\mu} - \frac{1}{2} \,(a_{\alpha} g_{\beta\mu} -a _{\beta} g_{\alpha\mu}),
\label{15111}
\end{equation}
where $ \sigma_{\mu\nu}$ is denoting the shear, and $\omega_{\alpha\beta}$ the vorticity, as defined in the Appendix. It is immediate to show that such tensor fulfills the conditions (\ref{17nov4})-(\ref{17nov5}) for a Lanczos tensor. Besides it satisfies the Lanczos gauge
$$ L_{\alpha\beta\mu} \, g^{\beta\mu} = 0.$$

In analogy with the Lorentz force (\ref{16novembro1}), the decomposition (\ref{15111}) suggests the unique form for the acceleration, that is,
\begin{equation}
a_{\alpha} =   L_{\alpha\beta\mu} \, v^{\beta} \, v^{\mu}.
 \label{16novembro2}
 \end{equation}
Note that this expression does not depends on the mass of the body. In other words, the gravitational field acts on H-poles in an universal way, regardless its mass. This is a direct consequence of the dimension $ lenght^{-1}$ of the Lanczos tensor. We note that in all these expressions we are setting the velocity of light $ c = 1.$ In the next section we present an explicit example of this formula.

\section{H-poles in the Schwarzschild metric}
Let us consider the geometry of a static spherically symmetric configuration given by

$$ ds^{2} = e^{\nu(r)} \, dt^{2} - e^{- \nu(r)}\, dr^{2} - r^{2} \, d\theta^{2} - r^{2} \, sin^{2}\theta \, d\varphi^{2} ,$$
where $ e^{\nu} = 1 - 2 m/r.$
In this case the gravitational force that drives the H-poles is controlled by the potential that obeys the equation
$$ \Box \Phi = 0.$$

The acceleration vector is given by
\begin{equation}
 a_{\mu} = \left(0, \frac{r_{H}}{2 r^{2} \, (1- r_{H}/r)}, 0, 0\right),
 \label{222}
 \end{equation}
that is, the acceleration is a gradient of $\Phi$,
$ a_{\mu} = \partial_{\mu} \Phi$, whose solution is
\begin{equation}
\Phi =  \frac{1}{2} \, \ln \left( 1 - \frac{r_{H}}{r} \right).
\label{241}
\end{equation}

\subsection{Effective metric}
According to General Relativity ordinary matter (which we denote by E-pole) follows geodesics in the gravitational metric $ g_{\mu\nu}.$ Let us now show that the gravitational magnetic pole (H-pole)that are accelerated by the gravitational field follows geodesics in an associated effective metric. In order to do this let us remind
the following Lemma \cite{novbit}:

\vspace{0.5cm}

\textbf{Lemma}

\vspace{0.70cm}
Consider an accelerated curve with velocity $ v^{\mu}$ in an arbitrary background metric $ g_{\mu\nu}$  such that its acceleration is a gradient  $a_{\mu} = \partial_{\mu} \Psi.$ It is always possible to construct an effective metric
\begin{equation}
 \hat{g}^{\mu\nu} =  g^{\mu\nu} +
\beta \, v^{\mu} \,v^{\nu},
\label{1802}
\end{equation}
such that this curve is mapped into a geodesics where
\begin{equation}
1 + \beta = e^{- 2 \Psi}.
  \label{18022}
  \end{equation}

The inverse covariant form of the effective metric is given by
 \begin{equation}
\hat{g}_{\mu\nu} = g_{\mu\nu} -
\frac{\beta}{ ( 1 + \beta)} \, v_{\mu} \,v_{\nu}.
\label{18023}
\end{equation}

In the case of an H-pole equation (\ref{11setembro2}) which implies that the acceleration on an H-pole is a gradient fulfils the condition of applicability of the above Lemma. Thus, it follows that H-poles follows geodesics in the metric (\ref{1802}) where $ \beta$ is related to the acceleration through condition (\ref{18022}).

\subsection{The equation of motion of an H-Pole}

Let $x^{\alpha}(s) $  be the path of an H-pole and its corresponding velocity $ (\dot{t}, \dot{r}, \dot{\theta}, \dot{\varphi})$. Its evolution is provided by
\begin{equation}
\frac{d}{ds} (g_{\alpha\lambda} \, \dot{x}^{\alpha}) - \frac{1}{2} \, g_{\mu\nu , \lambda} \, \dot{x}^{\mu} \, \dot{x}^{\nu} = \Phi_{, \lambda}.
\label{242}
\end{equation}
The comma means simple derivative: $ \Phi_{, \lambda} = \partial_{\lambda} \Phi$ and $ \Phi $ is given by (\ref{241}). From the equations of $ x^{2} = \theta$ and $ x^{3} = \varphi$ it follows that the angle $ \theta$ is a constant of motion. We choose $ \theta = \pi/2.$  For the angle $ \varphi$ we have
$$ \dot{\varphi} = \frac{h}{r^{2}} $$
for $ h $ constant. For the variable $ t $ it follows

$$ \dot{t} \,  (1 - \frac{r_{H}}{r}) = l$$
with $ l $ constant. From the auxiliary condition
$$ v^{\mu} \, v^{\nu} \, g_{\mu\nu} = 1 ,$$  we have
\begin{equation}
(1 - \frac{r_{H}}{r} ) \dot{t}^{2} - \frac{ \dot{r}^{2}}{(1 - r_{H}/r)} - \frac{h^{2}}{r^{2}} = 1.
\label{244}
\end{equation}
Once the acceleration has the form $ a_{\mu} = ( 0, a_{1}, 0, 0)$ it follows that $ dr/ds = 0.$ Thus the orbits of H-poles are only circular. That is
$$ v^{\mu} = (\frac{l}{(1 - r_H/R_{o})}, \, 0, 0, h/R_{o}^{2}).$$
The norm of $v^{\mu}$ implies
$$ l^{2} = ( 1 + \frac{h^2}{R_{o}^{2}}) \, ( 1 - r_{H}/R_{o})$$
where $ R_{o}$ is the constant radius of the circular orbit. We note that $ R_o > r_H.$

It then follows that the angular velocity of the H-pole is given by
$$ \omega^2 = \frac{r_H}{R_{o}^4} \, (R_o - r_H).$$

Let us note that the acceleration can be written under the formula of eq. (\ref{16novembro2}). Indeed, the velocity of the H-pole takes the form
$$ v^{\mu} = (\dot{t}, 0, 0, \dot{\varphi}).$$

The non-null components of Lanczos tensor in the Schwarzschild background reduces to $ L_{100} = r_{H}/2 \, r^{2}; L_{122} = L_{133} = - r_{H}/6(1- r_{H}/r). $
Thus it follows that we can write the acceleration eq. (\ref{222})  under the form of eq. (\ref{16novembro2}).

\vspace{0.3cm}

Just as it is extremely difficult to detect Dirac's magnetic monopole possibly the same could occur with gravitational magnetic monopoles. Nevertheless, further analysis on these hypothetical particles should be important to eventually elucidate the reason why Nature did not find it necessary to create them.

\section{Acknowledgement}
MN  would like to thank the support from brazilian agencies FAPERJ, CNPq and FINEP.

 \section{Appendix: Mathematical compendium}

Riemann curvature tensor can be decomposed into its irreducible parts by the relation

$$
R_{\alpha\beta\mu\nu} = W_{\alpha\beta\mu\nu} + M_{\alpha\beta\mu\nu} - \frac{1}{6} R g_{\alpha\beta\mu\nu}
$$
where $W_{\alpha\beta\mu\nu}$ is the Weyl conformal tensor,

$$
2 M_{\alpha\beta\mu\nu} = R_{\alpha\mu} g_{\beta\nu} + R_{\beta\nu} g_{\alpha\mu} - R_{\alpha\nu} g_{\beta\mu} - R_{\beta\mu} g_{\alpha\nu}
$$
and $ g_{\alpha\beta\mu\nu} = g_{\alpha\mu} \, g_{\beta\nu} - g_{\alpha\nu} \, g_{\beta\mu}.$
The dual operation for an arbitrary anti-symmetric tensor $ F_{\mu\nu} $ is defined by
$$ F^{*}_{\mu\nu} \equiv \frac{1}{2} \, \eta_{\mu\nu\alpha\beta} \, F^{\alpha\beta} $$
with
$$ \eta_{\alpha\beta\mu\nu} =  \sqrt{-g} \, \varepsilon_{\alpha\beta\mu\nu} .$$
$g$ is the determinant of  $g_{\mu\nu}$  and $\varepsilon_{\alpha\beta\mu\nu}$ is the Levi-Civita totally anti-symmetric quantity. We define the electric vector $ E^{\mu} $ and magnetic vector $ H^{\mu}$ by setting
$$E_{\alpha}=   F_{\alpha\mu} v^{\mu} ,$$
$$ H_{\alpha} =   F^{\ast}_{\alpha\mu} v^{\mu}.$$
The Weyl tensor has ten independent components and can also be separated by an arbitrary observer endowed with four velocity $v^{\mu}$ into its electric $ E_{\alpha\beta}$ and magnetic $ H_{\alpha\beta} $ tensor parts, that is
$$E_{\alpha\beta} =   W_{\alpha\mu\beta\nu} v^{\mu} v^{\nu},$$
$$ H_{\alpha\beta} =  W^{\ast}_{\alpha\mu\beta\nu} v^{\mu} v^{\nu}.$$
Thus electric and magnetic tensors are symmetric, traceless and orthogonal to the observer:
$$
E_{\mu\nu} = E_{\nu\mu}, \quad E_{\mu\nu} v^{\mu} = 0\quad {\rm and} \quad E_{\mu\nu} g^{\mu\nu} = 0
$$
$$
H_{\mu\nu} = H_{\nu\mu}, \quad H_{\mu\nu} v^{\mu} = 0 \quad {\rm and} \quad H_{\mu\nu} g^{\mu\nu} = 0.
$$

\subsection{The kinematical parameters }
%\vspace{0.50cm}
The shear $$ \sigma_{\mu\nu} = \frac{1}{2} h^{\alpha}{}_{( \mu} \, h^{\beta}{}_{\nu )} - \frac{1}{3} \, \theta \, h_{\mu\nu}$$
The vorticity $$  \omega_{\mu\nu} = \frac{1}{2} h^{\alpha}{}_{[ \mu} \, h^{\beta}{}_{\nu ]} $$
The expansion factor $$ \theta = v^{\alpha}{}_{; \alpha} $$
and the projection
$$ h_{\mu\nu} = \, g_{\mu\nu} - v_{\mu} \, v_{\nu}. $$
%Thus the shear is symmetric, trace-less and $ \sigma_{\mu\nu} \, v^{\mu} = 0.$ The vorticity is antisymmetric and $ \omega_{\mu\nu} \, v^{\mu} = 0.$


\newpage

\begin{thebibliography}{100}
	
\bibitem{novelloetal} M. Novello, C. A. P. Galv{\~a}o, I. Dami{\~a}o Soares and J. M. Salim, \textit{J. Phys. A Math. Gen.} \textbf{9}, 4 (1976) 547-554.
\bibitem{salim} J. M. Salim, Master Thesis, CBPF (1976, in portuguese).
\bibitem{dirac} P. A. M. Dirac, \textit{Phys. Rev.} \textbf{74}, 7 (1948) 817-830.

\bibitem{novellodamiao} M. Novello, I. Dami{\~a}o Soares and J. M. Salim, \textit{Gen. Rel. and Grav.} \textbf{8}, 2 (1977) 95-102.


\bibitem{lanczos} C. Lanczos, \textit{Rev. Mod. Phys.} \textbf{34}, 3 (1962) 379-389.
\bibitem{bampicaviglia} E. Bampi and G. Caviglia, \textit{Gen. Rel. Grav.} \textbf{15} (1983) 375-386.
\bibitem{novellovelloso} M. Novello and A. L. Velloso. \textit{Gen. Rel. Grav.} \textbf{19}, 12 (1987)
1251-1265.
\bibitem{bonilla} J. L. Lopez-Bonilla, G. Ovando, J. J. Pe{\~n}a, \textit{Found. Phys Lett} \textbf{12}, 4 (1999) 401-405.
\bibitem{light} We are setting the velocity of light in the natural units system $ c = 1.$
\bibitem{novbit} M. Novello and E. Bittencourt, \textit{Braz. J. of Phys.} \textbf{45} (2015) 756-805.	

\end{thebibliography}
 \end{document}